\documentclass{appolb}

\usepackage{amsmath,amssymb}
\usepackage{dsfont}

\newcommand{\be}{\begin{equation}}
\newcommand{\ee}{\end{equation}}
\newcommand{\ba}{\begin{eqnarray}}
\newcommand{\ea}{\end{eqnarray}}
\newcommand{\ban}{\begin{eqnarray*}}
\newcommand{\ean}{\end{eqnarray*}}

\begin{document}

\title{Enhancement of deuteron production in jets}

\author{Stanis\l aw Mr\' owczy\' nski
\address{National Centre for Nuclear Research, Warsaw, Poland \\ 
and \\
Institute of Physics, Jan Kochanowski University, Kielce, Poland}}

\maketitle

\begin{abstract}

A strong enhancement of deuteron production in jets has been recently observed in proton-proton collisions at LHC. We show that the effect is due to two independent factors: a collimation of jet nucleons and a smallness of nucleon source which is significantly smaller than a deuteron. 

\end{abstract}

Deuterons are abundantly produced in nucleus-nucleus collisions. At low-energy collisions they are fragments of incoming nuclei while at high energies we also deal with a genuine production process -- the energy released in a collision is converted into masses of mucleons which form deuterons. The remnants of initial nuclei occur at rapidities of the projectile and target while the genuinely produced deuterons populate a midrapidity domain. 

Recently, a production of deuterons associated with jets resulting from hard parton scatterings has been studied in proton-proton collisions at $\sqrt{s} = 13$ TeV at Large Hadron Collider (LHC) \cite{ALICE:2020hjy,ALICE:2022ugx}. Splitting the final state particles into those belonging to jets and those to the underlying events, it has been found that production of deuterons within jets is strongly enhanced. 

The experimental data have been compared with predictions of two variants of the coalescence model \cite{Butler:1963pp,Schwarzschild:1963zz}. The momentum distribution of nucleons has been provided by the Monte Carlo event generator  PYTHIA 8 \cite{Skands:2014pea} in both variants. In the first one a deuteron formation has been assumed to happen if a proton and neutron have a momentum difference below a coalescence momentum $p_0$ in the deuteron rest frame and the value of $p_0$ has been obtained fitting the data. In the second variant the deuteron formation probability is not either 0 or 1 but,  as proposed in \cite{Dal:2015sha}, it has been parameterized according to the energy dependent cross-sections of several reactions of deuteron production. Both variants of the coalescence model have described equally well the experimental data. The enhanced deuteron production in jets has been attributed to a reduced phase-space distance among jet nucleons. 

The aim of this note is to further discuss the enhanced production of jet deuterons. We show that there are actually two independent factors responsible for the effect. The first one is a collimated emission of jet nucleons and the second one is a smallness of a nucleon source which is significantly smaller than a deuteron. Both factors are discussed and quantitatively estimated using simple analytical formulas. The note had been prepared before the erratum of Ref.~\cite{ALICE:2022ugx} was published. The current version takes into account that the experimentally measured coalescence parameter ${\cal B}$, which enters Eq.~(\ref{cross-D-inv}), was reduced by factor of 3.

The coalescence model \cite{Butler:1963pp,Schwarzschild:1963zz} correctly describes deuteron production over a wide range of collision energies, see the review \cite{Mrowczynski:2020ugu} and recent studies \cite{Zhao:2022xkz,Hillmann:2021zgj,Mahlein:2023fmx,Florkowski:2023uim} with numerous references therein. The model is particularly reliable in case of the genuine production of deuterons as it happens in proton-proton collisions. Then, the process occurs in two steps: production of a pair of neutron and proton with small relative momentum and formation of a deuteron due to final-state interaction. Since the energy scale of the first step, which is a fourfold nucleon mass, is much bigger than that of the second one, which is a deuteron binding energy, a probability to produce a deuteron factorizes into the probability to produce the neutron-proton pair and the probability to form the deuteron. Therefore, the yield of deuterons with momentum ${\bf p}$ can be written in the factorized form 
\be
\label{cross-D}
\frac{dN_D}{d^3{\bf p}} = \frac{1}{2} {\cal A} \, 
\frac{dN_{np}}{d^3\big(\frac{1}{2}{\bf p}\big)d^3\big(\frac{1}{2}{\bf p}\big)} ,
\ee
where ${\cal A}$ is the deuteron formation rate discussed below and $\frac{dN_{np}}{d^3{\bf p}_n d^3{\bf p}_p}$ is the yield of pairs of neutron and proton with momenta ${\bf p}_n$ and ${\bf p}_p$. We have introduced the factor $1/2$ in Eq.~ (\ref{cross-D}) as the neutron-proton pair can be in two isospin states $I=1, I_3=0$ and $I=I_3=0$ while only the second one contributes to the deuteron production. The factor is compensated when the yield of neutron-proton pairs, which is usually not accessible experimentally, is expressed through the yield of proton-proton pairs (always in the state $I=I_3=1$) as   
\be
\frac{dN_{np}}{d^3{\bf p}_n d^3{\bf p}_p} = 2 \frac{dN_p}{d^3{\bf p}_n} \frac{dN_p}{d^3{\bf p}_p},
\ee
where the proton and neutron yields are assumed to be equal and independent of each other. Then, the formula (\ref{cross-D}) is rewritten as
\be
\label{cross-D-new}
\frac{dN_D}{d^3{\bf p}} = {\cal A} \, 
\bigg(\frac{dN_p}{d^3\big(\frac{1}{2}{\bf p}\big)} \bigg)^2 .
\ee

The deuteron formation rate is
\be
\label{d-form-rate}
{\cal A} = \frac{3}{4} (2\pi)^3 \int d^3r \, D ({\bf r})\, |\phi_d({\bf r})|^2 ,
\ee
where the source function $D ({\bf r})$ is a normalized to unity distribution of relative space-time positions of the neutron and proton at the kinetic freeze-out and $\phi _d({\bf r})$ is the deuteron wave function of relative motion. The factor $\frac{3}{4}$ reflects the fact the deuterons come from the neutron-proton pairs in the spin triplet state. It is obviously assumed that nucleons are unpolarized. The formula (\ref{d-form-rate}), which is written in the center-of-mass frame of the deuteron, does not assume, as one might think, that the two nucleons are emitted simultaneously. The vector ${\bf r}$ denotes the inter-nucleon separation at the moment when the second nucleon is emitted. For this reason, the function $D ({\bf r})$ effectively gives the space-time distribution. The formula (\ref{d-form-rate}), which is written in the natural units with $\hbar=c=1$, was discovered by Sato and Yazaki \cite{Sato:1981ez} and repeatedly discussed later on, see {\it e.g.} \cite{Mrowczynski:1987} and the review \cite{Mrowczynski:2020ugu}. 

A specific feature of a jet is a strong collimation of jet particles around the leading one. To show how it influences a deuteron production, we consider a highly simplified momentum distribution of jet nucleons. Using spherical coordinates with the $z$ axis along the trajectory of the leading particle, the yield of jet protons can be written as
\be
\label{distribution-N}
\frac{d N_p}{d^3{\bf p}} = N_p \frac{e^{-\alpha p}}{\pi \, \alpha^3} 
\frac{\Theta(\cos\theta - \cos\theta_c) }{1 - \cos\theta_c} ,
\ee
where $\alpha$ is a real positive parameter, $\Theta(x)$ is the step function and $\theta_c$ is the critical zenithal angle such that jet nucleons are produced within the cone of angle $\theta_c$. The angle expressed in radians approximately gives the so-called jet radius $R$. For a typical value of $R=0.2-0.5$ we have $\theta_c = 11^0 - 29^0$. The momentum distribution (\ref{distribution-N}) is normalized to $N_p$. 

Substituting the proton yield (\ref{distribution-N}) into Eq.~(\ref{cross-D-new}), the deuteron yield is
\be
\label{cross-D-model}
\frac{d N_D^{\rm jet}}{d^3{\bf p}} = {\cal A} \, N_p^2 \frac{e^{-\alpha p} }{\pi^2 \, \alpha^6} \, 
\frac{\Theta(\cos\theta - \cos\theta_c) }{(1 - \cos\theta_c)^2} ,
\ee
and the total yield of jet deuterons equals
\be
\label{cross-D-model-total}
N_D^{\rm jet} = \int d^3p \, \frac{d N_D^{\rm jet}}{d^3{\bf p}} 
=\frac{2{\cal A} \,N_p^2}{\pi \, \alpha^3 (1 - \cos\theta_c)} .
\ee
As one sees, the deuteron yield grows when $\theta_c$ decreases that is when the jet becomes more and more collimated. Reducing $\theta_c$ from $180^0$, when the momentum distribution is isotropic, to $\theta_c= 20^0$ the deuteron yield increases by the factor of 33. This result is obviously independent of the distribution of momentum $p$ which enters Eq.~(\ref{distribution-N}).

The formula (\ref{cross-D-model-total}) clearly shows that the enhancement of deuteron production due to the collimation of  nucleons is independent of the deuteron formation. So, let us discuss the formation rate ${\cal A}$ given by Eq.~(\ref{d-form-rate}). The source function $D({\bf r})$ is usually parameterized in the Gaussian form
\be
\label{Gauss}
D ({\bf r}) = \frac{e^{-\frac{{\bf r}^2}{4r_0^2}}}{(4 \pi r_0^2)^{3/2}} ,
\ee
where the parameter $r_0$ gives the root-mean-square of the nucleon source as $R_{\rm rms} =\sqrt{3}\,  r_0$. Using the parameterization (\ref{Gauss}) and the Hulth\'en wave function 
\be
\label{Hulthen}
\phi _d({\bf r}) = \sqrt{\frac{\alpha \beta (\alpha + \beta )}{2\pi (\alpha - \beta )^2}} \;
{e^{-\alpha r}-e^{-\beta r}  \over r } , 
\ee
with $\alpha = 0.23$ fm$^{-1}$ and $\beta = 1.61$ fm$^{-1}$ \cite{Hodgson:1971}, one finds numerically the formation rate ${\cal A}$ as a function of $r_0$, see Fig.~1 of Ref.~\cite{Mrowczynski:1992gc}. Since the wave function (\ref{Hulthen}) monotonously decreases as $r$ grows, the formation rate ${\cal A}$ is a monotonously decreasing function of $r_0$ with the maximum at $r_0 = 0$. The value of ${\cal A}$ at $r_0 = 0$ is reached when the radius of nucleon source is much smaller than the deuteron radius which is approximately 2 fm \cite{Babenko:2008zz}. 

One finds ${\cal A}$ at $r_0 = 0$ observing that  the function $|\phi _d({\bf r})|^2$ changes slowly around ${\bf r} = 0$ as compared to $D ({\bf r})$ when $r_0 \rightarrow 0$. Consequently, the value of $|\phi _d({\bf r})|^2$ at ${\bf r} = 0$ can be pulled out of the integral (\ref{d-form-rate}). Thus, one finds
\be
{\cal A}^{\rm max} \approx \frac{3}{4} (2\pi)^3 |\phi _d({\bf r}=0)|^2 \int d^3r \, D ({\bf r}) 
= \frac{3}{4} (2\pi)^3 |\phi _d({\bf r}=0)|^2.
\ee
The last equality holds because the source function is normalized to unity. Using the Hulth\'en wave function (\ref{Hulthen}), one finally finds  
\be
\label{d-form-rate-max}
{\cal A}^{\rm max} = 3\pi^2 \alpha \beta (\alpha + \beta ) \approx  20.2~{\rm fm}^{-3}.
\ee
The result is independent of the adopted parameterization of the source function and it fully depends on the wave function at ${\bf r}=0$. 

The coalescence parameter ${\cal B}$, which was measured in \cite{ALICE:2022ugx}, is defined analogously to ${\cal A}$ but the yields in Eq.~(\ref{cross-D-new}) are replaced by the Lorentz invariant yields that is
\be
\label{cross-D-inv}
E \frac{d N_D}{d^3{\bf p}} = {\cal B} \, 
\Big(\frac{E}{2} \frac{d N_p}{d^3\big(\frac{1}{2}{\bf p}\big)}\Big)^2 .
\ee

Since the formula (\ref{d-form-rate}) holds in the center-of-mass frame of the deuteron, one immediately finds that ${\cal B} = 2 {\cal A}/m$ where $m$ is the nucleon mass. The deuteron binding energy is ignored here. The maximal value of ${\cal B}$ corresponding to ${\cal A}^{\rm max}$ given in Eq.~(\ref{d-form-rate-max}) is ${\cal B}^{\rm max} \approx 0.33~{\rm GeV}^2$.

The experimentally found value of the coalescence parameter of jet deuterons is approximately ${\cal B} = 0.13 \pm 0.07~{\rm GeV}^2 $ \cite{ALICE:2022ugx} which is not far from the maximal value. Computing numerically the formation rate (\ref{d-form-rate}) or consulting Fig.~1 of Ref.~\cite{Mrowczynski:1992gc}, one finds the size of source of jet deuterons as $r_0 = 0.3 \pm 0.2~{\rm fm}$. We note that the coalescence parameter of deuterons in underlying events is about ten times smaller than that of jet deuterons \cite{ALICE:2022ugx} and $r_0 = 1.3 \pm 0.2~{\rm fm}$. 

Using a modern nucleon-nucleon potential like the Argonne v18 \cite{Wiringa:1994wb}, which takes into account a hard core repulsion at small distances, one finds that the deuteron wave function is significantly reduced at $r \lesssim 1$ fm when compared to the Hulth\'en wave function, see Fig. 11 of \cite{Wiringa:1994wb}. Consequently, the deuteron formation rate is significantly reduced, see Fig.~7 of \cite{Mahlein:2023fmx}, and it is too small to describe the deuteron production in jets. 

Here we arrive to a difficult problem of the relativistic deuteron structure, see the review \cite{Gilman:2001yh} and very recent publication \cite{Sargsian:2022rmq}. A potential such as Argonne is obtained from data on low-energy nucleon-nucleon scattering but the deuteron wave function at small distances corresponds to relativistic momenta. Then, a deuteron cannot be treated as a bound state of a proton and neutron but rather as a composite system of quarks and gluons. So, it is fair to say that the source of jet nucleons is much smaller than a deuteron and that the parameter $r_0$ is a fraction of 1 fm but further quantification of this conclusion is rather uncertain. 

The deuteron production requires a generation of at least six quark-antiquark pairs and their hadronization into nucleons. According to our conclusion, the whole process occurs at a length scale that is a fraction of a femtometer. Such a small length scale is not surprising if one takes into account that jets are produced in hard scatterings with a momentum transfer of at least a few GeV and of the same scale of offshellness of primarily produced partons. Consequently, the jet evolution occurs on a length scale of a fraction of a femtometer. 

At the end we note that the situation we have encountered that a particle's source is much smaller than a bound state formed by the particles emitted from the source is not exceptional. It occurs, for example, when hydrogen-like atoms of $\pi^+\pi^-$ are produced in high-energy proton-beryllium collisions \cite{DIRAC:2018xvz}, or in a Dalitz decay $\pi^0 \rightarrow \gamma (e^+ e^-)$ with the electron-positron pair in a positronium state \cite{Afanasev:1989ns}.  We also mention the decay $K^0_L \rightarrow (\pi^{\pm} \mu^{\mp}) \nu_\mu$ where the pion and muon are in a Coulomb bound state \cite{Aronson:1986qz}. In all these cases the bound-state formation is due to the final-state interaction and the cross section or decay branching ratio crucially depends on the bound-state wave function at ${\bf r}=0$. 

\vspace{3mm}

The useful correspondence from Maximilian Mahlein is gratefully acknowledged. This work was partially supported by the National Science Centre, Poland under grant 2018/29/B/ST2/00646. 

%-----------------------------------------------------------------

\end{document}